\documentclass[article,11pt,aps,prd,preprint,superscriptaddress,amsmath,amssymb,showpacs]{revtex4-1}
\usepackage{graphicx}
\usepackage{dcolumn}
\usepackage{bm}
\usepackage[colorlinks=true,allcolors=blue]{hyperref}
\usepackage{float}
\usepackage{physics}
\usepackage{slashed}
\usepackage{relsize}
\usepackage{amsmath}

\begin{document}
	
\title{The nucleon structure from an AdS/QCD model in the Veneziano limit}
	
\author{Jiali Deng}
\affiliation{CCNU, Wuhan, Inst. Part. Phys. and Hua-Zhong Normal U., LQLP}
\email{djl2022010355@mails.ccnu.edu.cn,
houdf@mail.ccnu.edu.cn
}	
\author{Defu Hou}
\affiliation{CCNU, Wuhan, Inst. Part. Phys. and Hua-Zhong Normal U., LQLP}

	
\date{\today}
\begin{abstract}
 We employ the VQCD model, a holographic approach that dynamically simulates essential QCD characteristics, including linear mass spectra, confinement, asymptotic freedom, and magnetic charge screening, while incorporating quark flavor effects. Using this model, we first calculate the proton mass spectrum and the wave function, incorporating anomalous dimensions to refine our results. Next, we compute the proton structure functions across a range of Bjorken $x$ values using consistent parameters. Furthermore, we derive the proton electromagnetic form factor by solving the electromagnetic field's motion equation, accounting for background effects, and demonstrate qualitative consistency with results from free electromagnetic fields coupled to fermions. Finally, we calculate the gravitational form factors by introducing an effective graviton mass $m$ arising from chiral symmetry breaking and the proton energy-momentum tensor.  Our calculations yield results that are in excellent agreement with experimental data and lattice QCD computations, validating the VQCD model as a robust tool for studying proton properties.
\end{abstract}
	
\maketitle	

\section{Introduction}\label{sec:01_intro}

The internal structure of nucleons remains one of the most intriguing puzzles in quantum chromodynamics (QCD). Understanding the proton's internal structure requires a detailed examination of its mass spectrum, structure functions, electromagnetic form factors, and gravitational form factors. These observables provide critical insights into the distribution of quarks and gluons, as well as the dynamics governing nucleon interactions. Previous studies, such as those in Refs\cite{Guidal:2013rya}, have extensively explored the physical meanings and interconnections of these quantities.

A significant challenge in studying nucleon structure arises from the strong coupling nature of QCD in the low-energy regime, where perturbative methods are inapplicable. To address this, the AdS/CFT duality has emerged as a powerful theoretical framework. This duality establishes a correspondence between a conformal field theory (CFT) in four-dimensional Minkowski spacetime and a string theory in five-dimensional anti-de Sitter (AdS) spacetime\cite{Maldacena:1997re,Witten:1998qj,Aharony:1999ti}. By breaking conformal invariance, phenomenological holographic models such as the soft wall model \cite{Karch:2006pv,He:2007juu} and the hard wall model \cite{Polchinski:2001tt,Boschi-Filho:2002xih} have been developed to bridge five-dimensional gravitational theories with QCD. These models have enabled the study of non-perturbative phenomena, including anomalous dimensions, which have been widely discussed in the literature \cite{Georgi:1974wnj,Jaroszewicz:1982gr}.

The proton mass spectrum is a key area of research, as it provides a testing ground for new methods to probe the proton's internal structure and compare theoretical predictions with experimental data. Significant progress has been made in calculating hadronic spectra using both the hard wall \cite{Erlich:2005qh,DaRold:2005mxj,deTeramond:2005su,Li:2013lfa} and soft wall models \cite{Vega:2008te,Branz:2010ub,Gutsche:2011vb,Fang:2016uer}. Additionally, modifications to the AdS metric, such as introducing a correction factor to the dilaton field, have been proposed to extend the applicability of these models to fermions\cite{Andreev:2006vy,Forkel:2007cm,Rinaldi:2017wdn,FolcoCapossoli:2019imm}.

Structure functions, which describe the parton distributions within the proton, offer valuable information about quark-gluon interactions. These functions are typically studied through deep inelastic scattering (DIS) experiments \cite{Polchinski:2002jw,BallonBayona:2007rs,BallonBayona:2007qr,Pire:2008zf,Taliotis:2009ne,Yoshida:2009dw,Cornalba:2009ax,Brower:2010wf,Koile:2014vca,Gao:2014nwa,FolcoCapossoli:2015hub,FolcoCapossoli:2020pks,Tahery:2021xsj}. Pioneering work in Ref\cite{Polchinski:2002jw} explored holographic DIS in the hard wall model, analyzing scalar and fermion contributions across different values of the Bjorken parameter $x$. Subsequent studies in Refs \cite{Tahery:2021xsj,FolcoCapossoli:2020pks} investigated DIS for baryons at small and large $x$, respectively, using a deformed $AdS_5$ metric. These studies revealed that the parameters of the structure functions vary significantly with $x$, with the best agreement to experimental data observed at larger $x$ values.

Electromagnetic form factors are fundamental observables that characterize the proton's charge and magnetic moment distributions\cite{Brodsky:1980zm,Pacetti:2014jai,Punjabi:2015bba}. For spin-1/2 particles, these are described by the Dirac and Pauli form factors, which can be extracted from elastic electron-nucleon scattering experiments \cite{Janssens:1965kd,Punjabi:2005wq,Puckett:2011xg}. Holographic models have been successfully applied to calculate these form factors, as demonstrated in Refs \cite{deTeramond:2006qmk,Grigoryan:2007vg,Grigoryan:2007my,Brodsky:2007hb,Abidin:2009hr,Brodsky:2014yha,Mondal:2016xpk,Sufian:2016hwn,Mamedov:2016ype,Mamo:2021jhj}.

Gravitational form factors (GFFs) encode essential information about the proton's mass, spin, and internal forces, providing insights into the so-called "proton mass and spin crisis" \cite{Brodsky:2000ii,Burkert:2023wzr,Nair:2024fit}. Analogous to electromagnetic form factors, GFFs describe interactions mediated by gravitons rather than photons. Calculations of GFFs for mesons and protons have been reported in Refs\cite{Brodsky:2008pf,Abidin:2008ku} and \cite{Abidin:2009hr,Hatta:2018ina,Mamo:2021krl}, respectively.

In the standard holographic QCD framework, the number of flavors $N_{f}$ is finite, while the number of colors $N_{C}$ approaches infinity, effectively neglecting quark effects. To address this limitation, Veneziano proposed an alternative limit \cite{Gursoy:2007cb,Gursoy:2007er,Jarvinen:2011qe,Arean:2013tja}:
\begin{equation}
	\label{eq1}
\ N_{C}\rightarrow \infty,   \quad N_{f}\rightarrow \infty,    \quad \frac{N_{f}}{N_{C}}=x_{f}.
\end{equation}
where $x_{f}$ is a fixed ratio. This approach, known as the holographic VQCD model, incorporates the running coupling constant and accounts for quark flavor effects.

Anomalous dimensions have been extensively studied within the holographic framework \cite{Tahery:2021xsj,FolcoCapossoli:2020pks,Vega:2008te,Boschi-Filho:2012ijd,FolcoCapossoli:2016uns,Rodrigues:2016kez}, particularly in the context of baryon and meson mass calculations \cite{Vega:2008te} and DIS studies \cite{Tahery:2021xsj,FolcoCapossoli:2020pks}. In this work, we employ the VQCD model to compute the proton's mass spectrum, structure functions, electromagnetic form factors, and gravitational form factors. Unlike conventional holographic models, the VQCD model features a dynamically determined correction factor in the $AdS_5$ metric, obtained by solving Einstein's equations. This deformation introduces a mass scale for proton fields while maintaining parameter consistency across physical processes. Our results demonstrate excellent agreement with experimental data and lattice calculations.

The paper is organized as follows: In Section II, we provide a brief overview of the VQCD model. Sections III through VI detail our calculations of the proton mass spectrum, structure functions, electromagnetic form factors, and gravitational form factors, respectively. Finally, Section VII summarizes our findings and discusses their implications.

\section{Holographic VQCD model}\label{sec:02}

The VQCD model consists of two parts: Improved holographic QCD model (IHQCD) and the tachyon DBI action \cite{Gursoy:2007cb,Gursoy:2007er,Jarvinen:2011qe,Arean:2013tja}. IHQCD corresponds to $\frac{N_{f}}{N_{C}}=0$ (i.e. a pure gluon system), and its action is
\begin{equation}
	\label{eq2}
S_{g}=M^{3}N_{C}^{2}\int d^{4}xdz\sqrt{-g}[R-\frac{4(\partial \lambda)^{2}}{3\lambda^{2}}+V_{g}(\lambda)],
\end{equation}
where $\lambda$  represents the running coupling constant and $V_{g}(\lambda)$ is the gluon potential function. The effect of quark flavor must be considered in finite $x_{f}$ , which is described by the DBI action:
\begin{equation}
	\label{eq3}
S_{f}=-x_{f}M^{3}N_{C}^{2}\int d^{4}xdz\sqrt{-g}V_{f}(\lambda, \tau)\sqrt{det(g_{\alpha \beta}+h(\lambda)\partial_{\alpha}\tau\partial_{\beta}\tau)},
\end{equation}
where the gauge field is zero, $V_{f}(\lambda,\tau)$  represents the flavor potential function and $\tau$ as the tachyon represents $D-\bar{D}$ string.

The action of the VQCD model is
\begin{equation}
	\label{eq4}
S=M^{3}N_{C}^{2}\int d^{4}xdz\sqrt{-g}[(R-\frac{4(\partial \lambda)^{2}}{3\lambda^{2}}+V_{g}(\lambda))-x_{f}V_{f}(\lambda, \tau)\sqrt{det(g_{\alpha \beta}+h(\lambda)\partial_{\alpha}\tau\partial_{\beta}\tau)}],
\end{equation}
Assuming the metric is
\begin{equation}
	\label{eq5}
ds^{2}=e^{2A}(-dt^{2}+d\vec{x}^{2}+dz^{2}).
\end{equation}

The gluon potential function $V_{g}(\lambda)$ can correspond to the Yang Mills field \cite{Gursoy:2007cb} at $x_{f}=0$. By comparison, the form of the gluon potential function can be written as
\begin{equation}
	\label{eq6}
V_{g}(\lambda)=12+\frac{44}{9\pi^{2}}\lambda+\frac{4619}{3888\pi^{4}}\frac{\lambda^{2}}{(1+\lambda /\lambda_{0})^{4/3}}\sqrt{1+Log(1+\lambda /\lambda_{0})}
\end{equation}
where $\lambda_{0}$ is a free parameter to avoid the occurrence of higher-order terms of potential energy in UV expansion. This form can qualitatively simulate some QCD characteristics such as linear mass spectrum, confinement, asymptotic freedom and magnetic charge screening \cite{Gursoy:2007er}.
The flavor potential function can compare QCD data ($x_{f}\neq0$)\cite{Gursoy:2007er}, the form of the flavor potential function can be written as
\begin{equation}
	\label{eq7}
V_{f0}(\lambda)=\frac{12}{11}+\frac{4(33-2x_{f})}{99\pi^{2}}+\frac{23473-2726x_{f}+92x_{f}^{2}}{42768\pi^{4}\lambda^{2}},
\end{equation}
\begin{equation}
	\label{eq8}
V_{f}(\lambda)=V_{f0}(\lambda)e^{-a(\lambda)\tau^{2}}, a(\lambda)=\frac{3}{22}(11-x_{f}), h(\lambda)=\frac{1}{(1+\frac{115-16x_{f}}{288\pi^{2}})^{4/3}}.
\end{equation}
To perform numerical solutions, we first select the infrared asymptotic expansion as $A(z)=-cz^{2}$ \cite{Gursoy:2007er}. Then, by solving Einstein's equations and equations of motion, infrared analytical solutions for other fields can be obtained. Finally, by substituting these boundary conditions, one can obtain the numerical solution of the VQCD model.

\section{Mass spectrum for proton}\label{sec:03}

In the quark model, baryon is a particle composed of three valence quarks. Baryon physics is introduced in reference \cite{Klempt:2009pi}. In this section, we will calculate the mass spectrum of proton.

In the holographic description, the proton is dual to a massive spinor field in $AdS_{5}$ space. The action of the spinor field can be written as\cite{FolcoCapossoli:2020pks}
\begin{equation}
	\label{eq9}
S=\int d^{4}xdz\sqrt{-g}\overline{\Psi}(\slashed{D}-m_{5})\Psi,
\end{equation}
where $m_{5}$ is the five-dimensional mass of a proton and the definition of covariant derivative is
\begin{equation}
	\label{eq10}
\slashed{D}=g^{mn}e^{a}_{n}(\partial_{m}+\frac{1}{2}\omega^{bc}_{m}\Sigma_{bc}),
\end{equation}
where the vierbein $e^{a}_{n}$ is determined by the metric, $\gamma_{a}=(\gamma_{\mu},\gamma_{5}),{\gamma_{a},\gamma_{b}}=2\eta_{ab}$ and $\Sigma_{bc}=\frac{1}{4}[\gamma_{b}, \gamma_{c}]$. $\gamma_{\mu}$ represent Dirac's gamma matrices and the indicators of flat space are represented by $a,b$ and $c$, Minkowski space is represented by $\mu,\nu$ and $m,n,p,q$ represent AdS space. The equation of motion is
\begin{equation}
	\label{eq11}
(\slashed{D}-m_{5})\Psi=0.
\end{equation}
Spin connection is defined as
\begin{equation}
	\label{eq12}
\omega^{ab}_{m}=e^{a}_{n}\partial_{m}e^{nb}+e^{a}_{n}e^{pb}\Gamma^{n}_{pm},
\end{equation}
with
\begin{equation}
	\label{eq13}
e^{a}_{n}=e^{A_{S}}\delta^{n}_{a},  e^{m}_{a}=e^{-A_{S}}\eta^{m}_{a},
\end{equation}
\begin{equation}
	\label{eq14}
\Gamma^{p_{mn}}=\frac{1}{2}g^{pq}(\partial_{n}g_{mq}+\partial_{mn}g_{nq}-\partial_{q}g_{mn}).
\end{equation}
then the equation of motion can be written as
\begin{equation}
	\label{eq15}
[e^{-A_{S}(z)}(\gamma^{5}\partial_{5}+\gamma^{\mu}\partial_{\mu}+2A'_{S}(z))-m_{5}]\Psi=0.
\end{equation}
Since spinor is either right-handed or left-handed, it can be decomposed into
\begin{equation}
	\label{eq16}
\Psi(x^{\mu,z})=[\frac{1+\gamma^{5}}{2}\chi_{R}(z)+\frac{1-\gamma^{5}}{2}\chi_{L}(z)]\Psi_{(4)}(x^{\mu}),
\end{equation}
where $\Psi_{(4)}(x^{\mu})$ is the four-dimensional wave function of a proton, which obeys the Dirac equation $(\slashed{D}-M)\Psi_{(4)}(x^{\mu})=0$ and since Kaluza-Klein modes correspond to chirality spinors, we can obtain
\begin{equation}
	\label{eq17}
\Psi_{R/L}(x^{\mu},z)=\underset{n}{\Sigma}\Psi_{R/L}^{n}(x^{\mu})\chi_{R/L}^{n}(z),
\end{equation}
where $\Psi_{R/L}^{n}(x^{\mu})=\frac{1\pm\gamma^{5}}{2}\Psi_{(4)}(x^{\mu})$ . By combining the equations of motion, one can obtain
\begin{equation}
	\label{eq18}
(\partial_{z}+2A'_{S}(z)+m_{5}e^{A_{S}(z)})\chi_{L}^{n}(z)=M_{n}\chi_{R}^{n}(z),
\end{equation}
\begin{equation}
	\label{eq19}
(\partial_{z}+2A'_{S}(z)-m_{5}e^{A_{S}(z)})\chi_{R}^{n}(z)=-M_{n}\chi_{L}^{n}(z).
\end{equation}
one introduces a transformation
\begin{equation}
	\label{eq20}
\chi_{R/L}^{n}(z)=e^{-2A_{S}(z)}\varphi_{R/L}^{n}(z).
\end{equation}
one can obtain a Schrödinger-like equation as
\begin{equation}
	\label{eq21}
-\varphi_{R/L}''(z)+[m_{5}^{2}e^{2A_{S}(z)}\pm m_{5}e^{A_{S}(z)}A_{S}'(z)]\varphi_{R/L}(z)=M_{n}^{2}\varphi_{R/L}(z),
\end{equation}
Where $M_{n}$ represents the four-dimensional proton mass. In pure AdS space, the five-dimensional mass can be written as
\begin{equation}
	\label{eq22}
m_{5}^{AdS}=|\triangle_{can}-2|.
\end{equation}
But QCD is not conformal invariant, it should be corrected by introducing anomalous dimensions. Then $m_{5}$ can be written as
\begin{equation}
	\label{eq23}
m_{5}=|\triangle_{can}-2|+\gamma.
\end{equation}
To calculate the mass spectrum of proton, we treat it as a particle (i.e. ignore its internal structure)\cite{FolcoCapossoli:2020pks} and take $\triangle_{can}=0.5$, which is the dimension of a fermion. The anomalous dimension $\gamma$ is fixed by the ground state mass of proton. Finally, by solving the Schrödinger-like equation, the excited state mass of the proton can be obtained. In this calculation, we selected constants: $c=0.25, \lambda_{0}=58\pi^{2}, m_{5}=0.279GeV$. Our results are very close to the experimental data, with an error within $3\%$.
\begin{table}
	\centering
	\includegraphics[width=8.5cm]{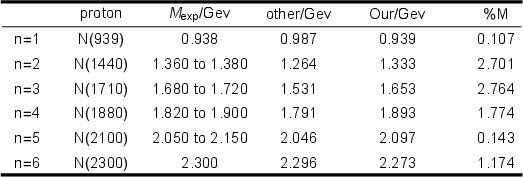}
	\caption{\label{Table 1}Masses of proton. The ground state mass is represented by $n=1$. $M_{exp}$ is experimental data from PDG \cite{ParticleDataGroup:2022pth}. Column Other is the result of the reference \cite{FolcoCapossoli:2019imm}. Column Our is our result and the last column represents the error.}
\end{table}

\section{Structure function of proton}\label{sec:04}

Electron scattering is a powerful tool to explore the internal structure of hadrons, as electrons have no internal structure and their interaction with target particles is mainly electromagnetic interaction, which can be accurately calculated using QED. Nucleon structure has been studied by two main types of electron scattering experiments: elastic scattering and deep inelastic scattering (DIS).
\begin{figure}
	\centering
	\includegraphics[width=8.5cm]{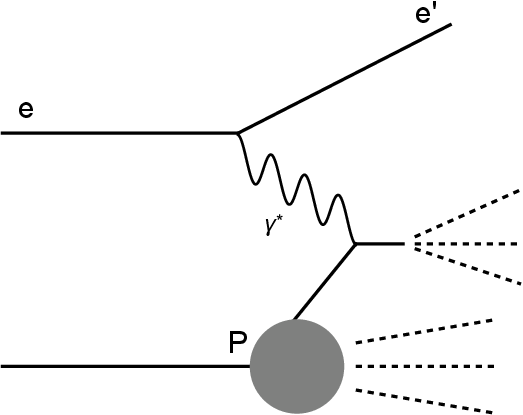}
	\caption{\label{figure}Deep inelastic scattering of a proton and
an electron by exchanging a virtual photon.}
\end{figure}

For DIS scattering, as shown in FIG.1. Electrons and protons interact by exchanging a virtual photon, where the squared four-momentum of the virtual photon is $Q^{2}$. Virtual photon interacts with individual parton of the proton, causing the parton to escape from the nucleon and undergo hadronization, leaving an undetermined final state.

To quantitatively calculate the DIS process, we will study scattering: $ep\rightarrow eX$, where X represents all possible final states. We can obtain the internal structure of the proton from fragmentation. The so-called Bjorken variable parametrizes this fragmentation according to:
\begin{equation}
	\label{eq24}
x=-\frac{q^{2}}{2pq},
\end{equation}
where p is the initial proton momentum and q is the transfer momentum from electron to proton through a virtual photon. The scattering amplitude can be written as:
\begin{equation}
	\label{eq25}
iM_{ep\rightarrow eX}=iQ\overline{u}\gamma^{\mu}u(\frac{i}{q^{2}})(ie)\int d^{4}xe^{iqx}\langle X|J^{\mu}(x)|p\rangle,
\end{equation}
where $J^{\mu}(x)$ represents quark electromagnetic current. By using the optical theorem, it is possible to obtain
\begin{equation}
	\label{eq26}
\underset{X}{\Sigma}d\Pi_{X}|M_{\gamma p\rightarrow X}|^{2}=2{\rm Im}M_{\gamma p\rightarrow \gamma p},
\end{equation}
in term of the spin-averaged forward matrix element between two proton currents, the hadron transition amplitude can be expressed as
\begin{equation}
	\label{eq27}
W^{\mu\nu}=\frac{i}{4\pi}\underset{s}{\Sigma}\int d^{4}xe^{iqx}\langle p,s|J^{\mu}(x)J^{\nu}(0)|p,s\rangle,
\end{equation}
where $|p,s\rangle$ represents the proton state with spin s. In Fourier space, we can obtain
\begin{equation}
	\label{eq28}
{\rm Im}W^{\mu\nu}=\frac{1}{4\pi}\underset{X}{\Sigma}\delta(M_{X}^{2}-(p+q)^{2})\langle p,s|J^{\mu}(0)|X\rangle \langle X|J^{\nu}(0)|p,s\rangle,
\end{equation}
where the final state must satisfy the energy conservation constraint.

The Ward-Takahashi identity requires that
\begin{equation}
	\label{eq29}
q_{\mu}W^{\mu\nu}=q_{\nu}W^{\mu\nu}=0.
\end{equation}
Therefore, the hadron tensor can be decomposed into
\begin{equation}
	\label{eq30}
W^{\mu\nu}=F_{1}(\eta_{\mu\nu}-q_{\mu}q_{\nu})+\frac{2x}{q^{2}}F_{2}(p^{\mu}+\frac{q^{\mu}}{2x})(p^{\nu}+\frac{q^{\nu}}{2x}),
\end{equation}
where $F_{1}$ and $F_{2}$ are the structure functions of protons.

Now we use the holographic VQCD model to calculate the structure function of protons. From the AdS/QCD dictionary, we can link the matrix element on the QCD side and the supergravity interaction action on the AdS side, $S_{int}$ .
\begin{equation}
	\label{eq31}
S_{int}=\eta_{\mu}\langle p+q,s_{X}|J^{\mu}(0)|p,s_{i}\rangle,
\end{equation}
where $\eta_{\mu}$ represents the polarization vector of the virtual photon, $s_{i}$ and $s_{X}$ are the spins of the initial proton and final hadron, respectively. In the AdS space, the interaction action can be written as
\begin{equation}
	\label{eq32}
S_{int}=g_{v}\int d^{4}xdz\sqrt{-g}\phi^{\mu}\overline{\Psi}_{X}\Gamma_{\mu}\Psi_{i},
\end{equation}
where $g_{v}$ is the coupling constant associated with the proton's electric charge, $\Gamma_{\mu}$ represent the gamma matrices of curved space. $\phi^{\mu}$ , $\Psi_{X}$ and $\Psi_{i}$ represent the electromagnetic field, initial proton field, and final hadron field, respectively.

The action of electromagnetic field can be written as
\begin{equation}
	\label{eq33}
S=-\frac{1}{4}\int d^{4}xdz\sqrt{-g}F^{mn}F_{mn},
\end{equation}
where $F_{mn}=\partial_{m}\phi_{n}-\partial_{n}\phi_{m}$. Then the equation of motion can be obtained
\begin{equation}
	\label{eq34}
\partial_{m}(\sqrt{-g}F^{mn})=0.
\end{equation}
Introduce the gauge conditions
\begin{equation}
	\label{eq35}
\partial_{\mu}\phi_{\mu}+e^{-A_{S}(z)}\partial_{z}(e^{A_{S}(z)}\phi_{z})=0.
\end{equation}
then the equation of motion becomes
\begin{equation}
	\label{eq36}
\partial_{\mu}\partial^{\mu}\phi_{\nu}+A'_{S}(z)\partial_{z}\phi_{\nu}+\partial_{z}^{2}\phi_{\nu}=0,
\end{equation}
\begin{equation}
	\label{eq37}
\partial_{\mu}\partial^{\mu}\phi_{z}+\partial_{\mu}\partial_{z}\phi_{z}=0.
\end{equation}

For numerical solution, we assume $\phi_{\nu}=\eta_{\nu}e^{iqx}\phi(z)$. At the boundary , we take $\phi(0)=1$ and take the Neumann boundary condition at the IR limit.

Based on the analysis in the previous chapter, the motion equation satisfied by the spinor field is
\begin{equation}
	\label{eq38}
-\varphi_{R/L}''(z)+[m_{5}^{2}e^{2A_{S}(z)}\pm m_{5}e^{A_{S}(z)}A_{S}'(z)]\varphi_{R/L}(z)=M_{n}^{2}\varphi_{R/L}(z),
\end{equation}
The initial and final state wave functions can be written as
\begin{equation}
	\label{eq39}
\Psi_{i}=e^{ipx}e^{-2A_{S}(z)}[\frac{1+\gamma^{5}}{2}\varphi_{R}(z)+\frac{1-\gamma^{5}}{2}\varphi_{L}(z)]u_{s_{i}(p)},
\end{equation}
\begin{equation}
	\label{eq40}
\Psi_{X}=e^{ip_{X}x}e^{-2A_{S}(z)}[\frac{1+\gamma^{5}}{2}\varphi_{R}(z)+\frac{1-\gamma^{5}}{2}\varphi_{L}(z)]u_{s_{X}(p_{X})}.
\end{equation}
According to the above calculation, the interaction quantity can be obtained
\begin{equation}
	\label{eq41}
S_{int}=g_{v}\int d^{4}xdz\sqrt{-g}g^{\mu\nu}\phi_{\nu}\overline{\Psi}_{X}e^{A_{S}}\delta_{\mu}^{m}\gamma_{m}\Psi_{i}=g_{v}\int d^{4}xdze^{4A_{S}}\eta^{\mu\nu}\phi_{\nu}\overline{\Psi}_{X}\gamma_{\mu}\Psi_{i}
\end{equation}
and from (40), we get
\begin{equation}
	\label{eq42}
\overline{\Psi}_{X}=e^{-ip_{X}x}e^{-2A_{S}(z)}\overline{u}_{s_{X}}(p_{X})[\frac{1-\gamma^{5}}{2}\varphi_{R}(z)+\frac{1+\gamma^{5}}{2}\varphi_{L}(z)].
\end{equation}
Therefore, (40) can be written as
\begin{equation}
\begin{split}
	\label{eq43}
S_{int}&=\frac{g_{v}}{2}\int d^{4}xdz[\overline{u}_{s_{X}}(\frac{1-\gamma^{5}}{2}\varphi_{R}(z)+\frac{1+\gamma^{5}}{2}\varphi_{L}(z))\gamma_{\mu}(\frac{1+\gamma^{5}}{2}\varphi_{R}(z)+\frac{1-\gamma^{5}}{2}\varphi_{L}(z))]u_{s_{i}}\\
       &=\frac{g_{v}}{2}(2\pi)^{4}\delta^{4}(p_{X}-p-q)\eta_{\nu}[\overline{u}_{s_{X}}\gamma^{\mu}\widehat{P}_{L}u_{s_{i}}I_{L}+\overline{u}_{s_{X}}\gamma^{\mu}\widehat{P}_{R}u_{s_{i}}I_{R}]\\
\end{split}
\end{equation}
with
\begin{equation}
	\label{eq44}
I_{R/L}=\int dz\phi(z)\varphi_{R/L}^{X}(z)\varphi_{R/L}^{i}(z).
\end{equation}
and (31) can be written as
\begin{equation}
	\label{eq45}
\eta_{\mu}\langle p_{X},s_{X}|J^{\mu}(0)|p,s_{i}\rangle=\frac{g_{eff}}{2}\delta^{4}(p_{X}-p-q)\eta_{\mu}[\overline{u}_{s_{X}}\gamma^{\mu}\widehat{P}_{L}u_{s_{i}}I_{L}+\overline{u}_{s_{X}}\gamma^{\mu}\widehat{P}_{R}u_{s_{i}}I_{R}],
\end{equation}
\begin{equation}
	\label{eq46}
\eta_{\mu}\langle p,s_{i}|J^{\nu}(0)|p_{X},s_{X}\rangle=\frac{g_{eff}}{2}\delta^{4}(p_{X}-p-q)\eta_{\nu}[\overline{u}_{s_{i}}\gamma^{\nu}\widehat{P}_{L}u_{s_{X}}I_{L}+\overline{u}_{s_{i}}\gamma^{\nu}\widehat{P}_{R}u_{s_{X}}I_{R}].
\end{equation}
Then the hadron tensor can be written as
\begin{equation}
\begin{split}
	\label{eq47}
\eta_{\mu}\eta_{\nu}W^{\mu\nu}&=\frac{\eta_{\mu}\eta_{\nu}}{4}\underset{M_{X}^{2}}{\Sigma}\underset{s_{i},s_{X}}{\Sigma}\frac{g_{eff}^{2}}{4}\delta(M_{X}^{2}-(p+q)^{2})
(\overline{u}_{s_{X}}\gamma^{\mu}\widehat{P}_{L}u_{s_{i}}\overline{u}_{s_{i}}\gamma^{\nu}\widehat{P}_{L}u_{s_{X}}I_{L}^{2}\\
&+\overline{u}_{s_{X}}\gamma^{\mu}\widehat{P}_{R}u_{s_{i}}\overline{u}_{s_{i}}\gamma^{\nu}\widehat{P}_{R}u_{s_{X}}I_{R}^{2}
+\overline{u}_{s_{X}}\gamma^{\mu}\widehat{P}_{L}u_{s_{i}}\overline{u}_{s_{i}}\gamma^{\nu}\widehat{P}_{R}u_{s_{X}}I_{R}I_{L}\\
&+\overline{u}_{s_{X}}\gamma^{\mu}\widehat{P}_{R}u_{s_{i}}\overline{u}_{s_{i}}\gamma^{\nu}\widehat{P}_{L}u_{s_{X}}I_{R}I_{L}).
\end{split}
\end{equation}
by summing up the spins, we can obtain
\begin{equation}
	\label{eq48}
\underset{s_{i}}{\Sigma}(u_{s_{i}})_{\alpha}(p)(\bar{u}_{s_{i}})_{\beta}(p)=(\gamma^{\mu}p_{\mu}+M)_{\alpha\beta}
\end{equation}
Therefore, the hadron tensor can be written as
\begin{equation}
	\label{eq49}
\eta_{\mu}\eta_{\nu}W^{\mu \nu}=\frac{1}{4}\underset{M_{X}^{2}}{\Sigma}\delta(M_{X}^{2}-(p+q)^{2}){[(p_{X}\cdot\eta)^{2}-\frac{1}{2}\eta\cdot\eta (p_{X}^{2}+p_{X}\cdot q)](I_{R}^{2}+I_{L}^{2})+M_{X}M\eta\cdot\eta. I_{R}I_{L}}
\end{equation}
To obtain the structural function, the delta function takes the following form\cite{Polchinski:2002jw,BallonBayona:2007rs}
\begin{equation}
	\label{eq50}
\delta(M_{X}^{2}-(p+q)^{2})\propto(\frac{\partial M_{n}^{2}}{\partial n})^{-1}\sim(2\pi s^{1/2}\Lambda)^{-1}.
\end{equation}
Considering transversal polarization, the structure function of protons can be written as
\begin{equation}
	\label{eq51}
F_{2}(x,q^{2})=\frac{g_{eff}^{2}}{8M_{X}}\frac{q^{2}}{x}(I_{R}^{2}+I_{L}^{2}).
\end{equation}

Now we numerically calculate the proton structure function for some fixed Bjorken variables $x=0.56, 0.65, 0.75$ and $0.85$, which are the highest values corresponding to the experiments. Firstly, we obtain the five-dimensional mass $m_{5}=0.229GeV$ based on the ground state mass of proton by numerically solving equation.(38). The ground state (n=1) represents the wave function of the target proton, as shown in Figure 2, while other states (n=2, 3...) are possible final state wave functions. The effective coupling constants are obtained by fitting the proton structure function, and the corresponding coupling constants are 2.92, 2.42, 1.66, and 1.08 for $x$ from small to large. Here, the five-dimensional mass of the proton we selected is different from the calculated mass spectrum, as the final state of DIS is not the excited state of the proton.
\begin{figure}
	\centering
	\includegraphics[width=8.5cm]{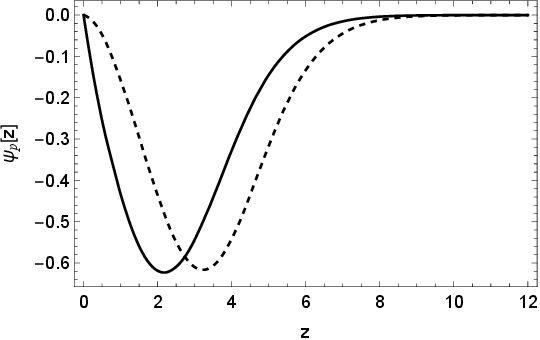}
	\caption{\label{figure}Chiral wave functions of the target proton with $m_{5}=0.229GeV$. The dashed and solid lines represent the left-handed and right-handed wave functions, respectively
.}
\end{figure}
\begin{figure}
	\centering
	\includegraphics[width=8.5cm]{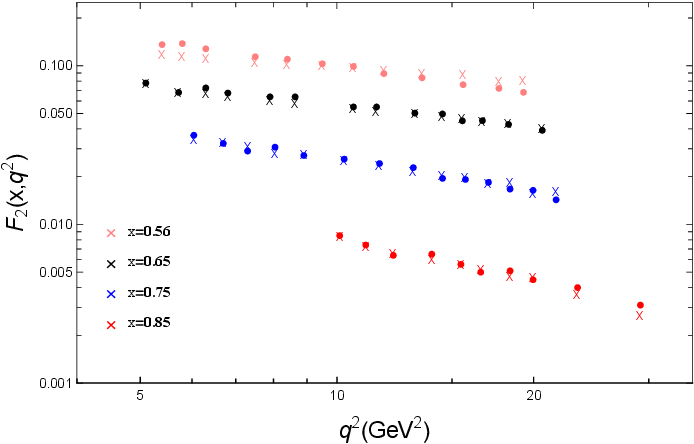}
	\caption{\label{figure}Comparison between our theoretical results and SLAC data.
The fork represents the experimental result \cite{ParticleDataGroup:2022pth}, and the dots are our results
.}
\end{figure}

Figure 3 shows the relationship between proton structure function and the square of the transfer momentum. Fork represents the experimental result, and the dots are our results, corresponding to $x=0.56, 0.65, 0.75$ and $0.85$ from top to bottom, respectively. It can be observed that our results are quite close to the experimental data.

Compared with reference.\cite{FolcoCapossoli:2020pks}, in our calculation, only the effective coupling constant is different, while other parameters are consistent for different $x$. Since virtual photons mainly interact with individual parton within proton in DIS, the effective coupling constant depends on the parton involved in the interaction. In addition, when $x=0.85, 0.75$, most of the momentum is transferred to a single individual parton rather than exciting the entire hadron, so the final state may still be the ground state. As $x$ decreases, the momentum carried by the parton involved in the interaction decreases, leading to an increase in the energy of the final state system. Therefore, the final state is usually not the ground state. In the calculation, we excluded the possibility of the final state being the ground state for $x=0.65, 0.56$. The structure function we calculated is consistent with the experimental data.

\section{Electromagnetic form factors}\label{sec:05}

For elastic scattering, as shown in Figure 3. Electrons and protons interact by exchanging a virtual photon, where the squared four-momentum of the virtual photon is $Q^{2}$. The interaction between virtual photon and individual parton inside the nucleon is different from the DIS process, as the parton still remains in the nucleon. Therefore, after the interaction, the final state remains a nucleus, with only the momentum changing.
\begin{figure}
	\centering
	\includegraphics[width=8.5cm]{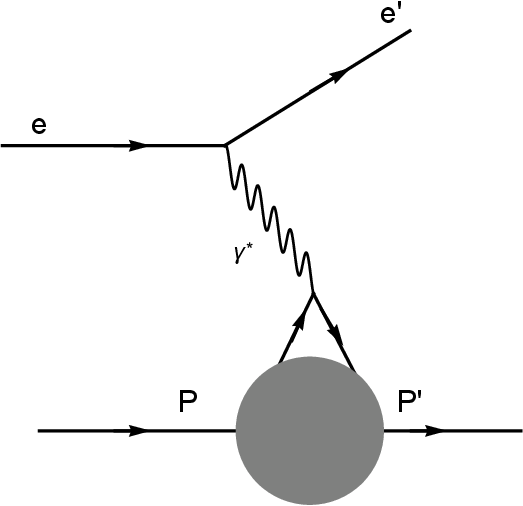}
	\caption{\label{figure}Elastic scattering of a proton and an electron
by exchanging a virtual photon
.}
\end{figure}

This elastic scattering can be described by two form factors:
\begin{equation}
	\label{eq52}
\langle p'|J^{\mu}(0)|p\rangle=\bar{u}(p')[\gamma^{\mu}F_{1}(q^{2})+\frac{i\sigma^{\mu\nu}q_{\nu}}{2M}F_{2}(q^{2})]u(p),
\end{equation}
where $\sigma^{\mu\nu}=[\gamma^{\mu},\gamma^{\nu}]$ and $q=p'-p$ is the four-momentum of a virtual photon. $F_{1}(q^{2})$ is the Dirac form factor, corresponding to the spin conserved flow matrix element, and $F_{2}(q^{2})$ is the Pauli form factor, corresponding to the spin flipped flow matrix element. They are respectively related to the distribution of charges and magnetic moments inside the nucleon.

In AdS space, the spin conserved flow matrix elements can be represented as
\begin{equation}
	\label{eq53}
\int d^{4}xdz\sqrt{-g}\bar{\Psi}_{p'}(x,z)e^{M}_{A}\Gamma^{A}\phi_{M}(x,z)\Psi_{p}(x,z)\sim (2\pi)^{4}\delta^{4}(p'-p-q)\epsilon_{\mu}\bar{u}(p')\gamma^{\mu}F_{1}(q^{2})u(p),
\end{equation}
where $\Gamma^{A}=(\gamma^{\mu},\gamma^{5})$ and the $e^{M}_{A}$ represent the inverse vierbein. The Dirac form factor can be expressed as
\begin{equation}
	\label{eq54}
F_{1}(q^{2})=\underset{R/L}{\Sigma}g_{R/L}^{N}\int dze^{4A_{S}(z)}\phi(z)\chi_{R/L}^{2}(z).
\end{equation}
The spin flipped flow matrix element can be written as
\begin{equation}
	\label{eq55}
\int d^{4}xdz\sqrt{-g}\bar{\Psi}_{p'}(x,z)e^{M}_{A}e^{N}_{B}[\Gamma^{A},\Gamma^{B}]F_{MN}(x,z)\Psi_{p}(x,z)\sim (2\pi)^{4}\delta^{4}(p'-p-q)\epsilon_{\mu}\bar{u}(p')\frac{\sigma^{\mu\nu}q_{\nu}}{2M}F_{2}(q^{2})u(p),
\end{equation}
The Dirac form factor can be expressed as
\begin{equation}
	\label{eq56}
F_{2}(q^{2})=\eta_{N}\int dze^{3A_{S}(z)}\phi(z)\chi_{R}(z)\chi_{L}(z),
\end{equation}
where $N=p,n.$ The effective charge $g_{R/L}^{N}$ cannot be calculated using holographic principles and can only be determined through specific spin-flavor structure. In the SU (6) approximation, the effective charge can be obtained as
\begin{equation}
	\label{eq57}
g_{R}^{p}=1, g_{L}^{p}=0, g_{R}^{n}=-\frac{1}{3}, g_{L}^{n}=\frac{1}{3}.
\end{equation}
Through comparative experiments, we can conclude that $\eta_{p}=1.793,\eta_{n}=-1.913$.

The action of the electromagnetic field can be written as
\begin{equation}
\begin{split}
	\label{eq58}
S_{f}&=-xM^{3}N_{C}^{2}\int d^{4}xdz\sqrt{-g}V_{f}(\lambda, \tau)(\sqrt{det(g_{\alpha \beta}+h(\lambda)\partial_{\alpha}\partial_{\beta}\tau+h(\lambda)F_{\alpha\beta}^{L})}\\
&+\sqrt{det(g_{\alpha \beta}+h(\lambda)\partial_{\alpha}\partial_{\beta}\tau+h(\lambda)F_{\alpha\beta}^{R})})
\end{split}
\end{equation}
The equation of motion is
\begin{equation}
	\label{eq59}
\frac{1}{V_{f}(\lambda, \tau)h(\lambda)^{2}e^{A}G}\partial_{z}(V_{f}(\lambda, \tau)h(\lambda)^{2}e^{A}G^{-1}\partial_{z}\phi(z))-Q^{2}\phi(z)=0,
\end{equation}
where $G=\sqrt{1+e^{-2A}h(\lambda)\tau'(z)^{2}}$.
\begin{figure}
	\centering
	\includegraphics[width=8.5cm]{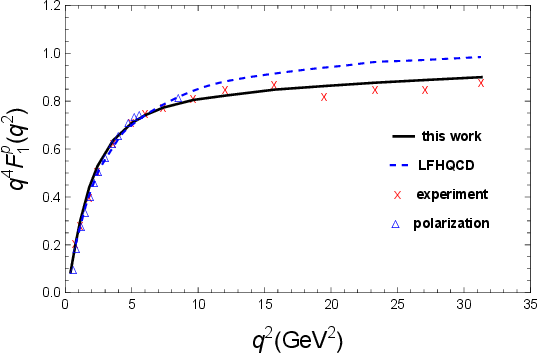}
	\caption{\label{figure}Comparison of Dirac form factor our result
with experimental data \cite{Riordan:2010id,Qattan:2012zf} and other model results\cite{Brodsky:2014yha}
.}
\end{figure}

Fig. 5 shows the relationship between proton Dirac form factors and the square of the transfer momentum. Fork and triangle represent the experimental result, the dashed line is the result of light-front holographic and solid line represent our result. We can see that both our model and light-front holographic model agree well with low transfer momentum, but as the transfer momentum increases, light-front holographic model deviates significantly from the experimental data, and our results agree better with the experiment. Fig. 6 shows the relationship between proton Pauli form factors and the square of the transfer momentum. We can also see that our results are more consistent with experimental data than those of light-front holographic model.
\begin{figure}
	\centering
	\includegraphics[width=8.5cm]{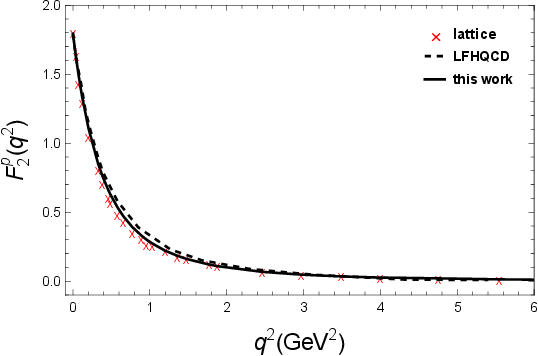}
	\caption{\label{figure}Comparison of Pauli form factor our result
with experimental data \cite{Riordan:2010id,Qattan:2012zf} and other model results\cite{Brodsky:2014yha}
.}
\end{figure}

Our model considers the effect of quark flavor. Firstly, the background metric is obtained directly from the Einstein equations and equations of motion corresponding to the non-gauge field action. Then, a gauge field is introduced into the DBI action, and the solution of the gauge field is obtained by solving its motion equation. Finally, substitute to obtain the electromagnetic form factor of the proton. That is to say, the interaction between electromagnetic field and flavor has been considered. From Figure 5, the results of light-front holographic model show a significant increase compared to the experimental data, which is qualitatively consistent with our model's consideration of free electromagnetic fields. And our results are consistent with considering the coupling of free electromagnetic fields and fermions. This is because in the low-energy region, the form factor of proton is mainly determined by their overall charge and magnetic moment distribution. Within this energy range, protons can be viewed as a whole, and the influence of their internal structural details is relatively small. In the high-energy region, the internal structure of proton becomes more prominent, and the role of flavors becomes more apparent.

\section{Gravitational form factors}\label{sec:06}

The origin of proton masses has been a significant mystery since the early days of QCD. One might ask, at the most basic level, how the QCD action, which involves light quarks and massless gluons, accounts for the proton's mass of about 1 GeV? Another key question pertains to the proton's spin. One would initially assume that the proton's spin of 1/2 comes from the quark’s spins, in a 'static' quark model. However, the precise decomposition of the proton's spin remains unknown at the quantitative level. To address these issues, it is necessary to study the energy-momentum tensor matrix elements, which encode information about proton spin, mass and internal pressure.

The matrix elements of the proton's energy tensor can be written as
\begin{equation}
	\label{eq60}
\langle p'|T^{\mu\nu}(0)|p\rangle=\bar{u}(p')[\gamma^{(\mu}p^{\nu)}A(q^{2})+\frac{ip^{(\mu}\sigma^{\nu)\alpha}q_{\alpha}}{2M}B(q^{2})+\frac{q^{\mu}q^{\nu}-\eta^{\mu\nu}q^{2}}{2M}D(q^{2})]u(p),
\end{equation}

The energy tensor of a proton is described by three form factors: $A(q^{2})$ describes the momentum distribution of the proton, $B(q^{2})$ is related to the angular momentum distribution of the proton, and $D(q^{2})$ is related to the internal pressure of the proton. The energy tensor can be expressed as
\begin{equation}
	\label{eq61}
T^{\mu\nu}(x)=-\frac{2}{\sqrt{-g}}\frac{\delta S_{M}}{\delta g^{\mu\nu}(x)},
\end{equation}
here $S_{M}$ represents the energy tensor of proton, which causes perturbations in the background spacetime: $\bar{g}_{mn}=g_{mn}+h_{mn}$ . Then expand $S_{M}$ to obtain
\begin{equation}
	\label{eq62}
S_{M}(h_{mn})=S_{M}(0)+\frac{1}{2}\int d^{5}x\sqrt{-g}h_{mn}T^{mn}.
\end{equation}
Substitute the metric perturbation into the gravitational force and select the gauge conditions. The equation of motion for graviton can be expressed as
\begin{equation}
	\label{eq63}
\frac{1}{\sqrt{-g}}\partial_{M}(\sqrt{-g}g^{mn}\partial_{N}h_{\mu\nu})+m^{2}h_{\mu\nu}=0.
\end{equation}
The $h_{\mu\nu}$ here can be considered as a graviton, which can be written as
\begin{equation}
	\label{eq64}
h_{\mu\nu}(x,z)=\varepsilon_{\mu\nu}e^{-iqx}H(q^{2},z).
\end{equation}

It satisfies the boundary condition $H(q^{2},0)=1$ and is zero at infinity. By substituting them into the equation of motion, we can obtain the numerical solution of the graviton. In standard theory, the mass of a graviton is zero. In our model, the background spacetime is not a pure AdS spacetime, it exists in a matter field (i.e. quark flavor and proton action), so we introduce an effective mass m, which is a free parameter.

The energy tensor of a proton can be written as
\begin{equation}
	\label{eq65}
T^{\mu\nu}(x,z)=\frac{i}{2}e^{-A_{S}(z)}\bar{\Psi}\gamma^{\mu}\partial^{\nu}\Psi-g^{\mu\nu}\ell.
\end{equation}
The five-dimensional interaction action can be expressed as
\begin{equation}
	\label{eq66}
S=\int d^{5}x\sqrt{-g}(\frac{-ie^{-A_{S}(z)}h_{\mu\nu}(x,z)}{4})\bar{\Psi}\gamma^{\mu}\partial^{\nu}\Psi.
\end{equation}
Comparing equations 60 and 66, we can obtain the form factor:
\begin{equation}
	\label{eq67}
A(Q^{2})=\int dzH(q^{2},z)(\varphi_{L}^{2}(z)+\varphi_{R}^{2}(z)).
\end{equation}
\begin{figure}
	\centering
	\includegraphics[width=8.5cm]{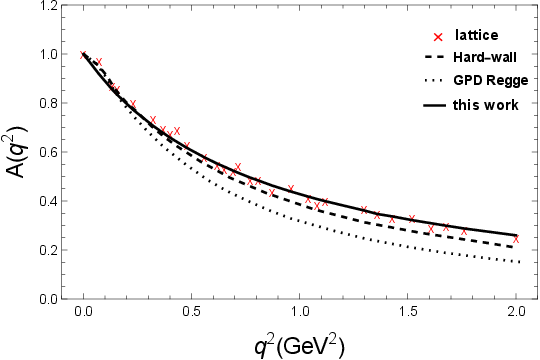}
	\caption{\label{figure}Comparison of gravitational form factor $A(Q^{2})$ theory
results with lattice \cite{Hackett:2023rif} data and other model results\cite{Abidin:2009hr}
.}
\end{figure}

Fig. 7 shows the relationship between proton gravitational form factor $A(Q^{2})$ and the square
of the transfer momentum. Fork represents the result of lattice calculation, the dashed line is the GPD model, the dotted line is the hard-wall model and solid line is our result with $m^{2}=0.02$. We can see that our results are more consistent with the lattice results.

The Pauli-like form factor $B(Q^{2})$ can be expressed as
\begin{equation}
	\label{eq68}
B(Q^{2})=\int dze^{-A_{S}(z)}H(q^{2},z)(\varphi_{L}(z)\varphi_{R}(z)).
\end{equation}
Fig. 7 shows the relationship between proton gravitational form factor $A(Q^{2})$ and the square
of the transfer momentum. Fork represents the result of lattice calculation and solid line is our result with $m^{2}=0.08$. Our results are consistent with the lattice results.
\begin{figure}
	\centering
	\includegraphics[width=8.5cm]{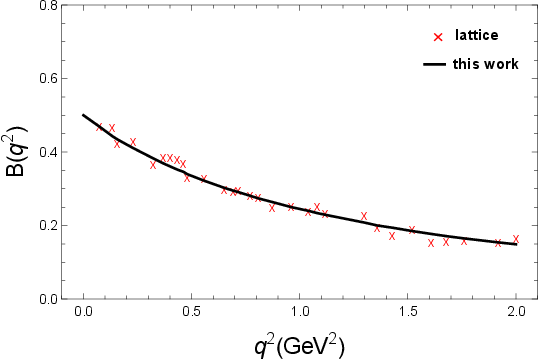}
	\caption{\label{figure}Comparison of gravitational form factor $B(Q^{2})$ theory
results with lattice data \cite{Hackett:2023rif}
.}
\end{figure}

In this calculation, we selected parameter $c=0.25,\lambda_{0}=58\pi^{2},m_{5}=0.279GeV$, which is the same as when calculating the proton mass spectrum, because its interaction is relatively weak, and the final state is still the proton ground state. The equation of motion for graviton is zero on the right-hand side, but this is incomplete. We should consider the influence of the source (i.e. the proton's energy tensor), although it is relatively small. Since each component of the proton's energy tensor is different, the duality formula cannot be directly applied. Assuming the graviton mainly relies on the zero component of the proton energy tensor for the gravitational form factor $A(Q^{2})$, while the gravitational form factor $B(Q^{2})$ mainly depends on the space component of the proton energy tensor. This is also the reason why the two form factors correspond to different effective masses $m$. We can find that the calculation results are consistent with our introduction of effective mass qualitatively.

\section{Summary and conclusions}\label{sec:07}

We investigated the nucleon structure using the holographic VQCD model, which offers a closer approximation to QCD compared to the soft wall model of pure AdS metric and the deformed AdS metric holographic model. This model dynamically simulates key QCD characteristics, including linear mass spectra, confinement, asymptotic freedom, magnetic charge screening, and quark flavor effects. Additionally, we incorporated the anomalous dimension of mass due to chiral symmetry breaking to study the nuclear structure.

 Our primary achievements include a comprehensive analysis of the proton structure by calculating its mass spectrum, structure function, electromagnetic form factor, and gravitational form factor.
 The mass spectrum, obtained by solving a Schrödinger-like equation with the inclusion of the anomalous dimension, shows remarkable agreement with experimental data, with errors within $3\%$.

 The calculation of the proton's structure function further demonstrates the model's capability to capture the dynamics of parton interactions. The variation in the effective coupling constant $g_{eff}$ with different longitudinal momentum fractions $x$ reflects the changing interaction strength between the virtual photon and the parton. Our results also exclude the possibility of the final state being the ground state at specific $x$ values, highlighting the increasing energy of the final state system as $x$ decreases. This finding provides deeper insights into the parton distribution within the proton.

In the calculation of the electromagnetic form factors, our approach of directly introducing the electromagnetic field into the background action and considering quark flavor interactions yields results that are in closer agreement with experimental data, particularly at high energies. This is attributed to the model's ability to account for the internal structure of the proton, which becomes more significant at higher energies. In contrast, at lower energies, the proton's form factor is predominantly influenced by its overall charge and magnetic moment distribution, making the internal structure less relevant.

Finally, the gravitational form factor calculation, which incorporates the effective mass derived from chiral symmetry breaking and the energy tensor of the proton, shows better agreement with lattice calculations compared to hard wall models. This consistency is achieved by qualitatively considering the energy tensor in the motion equation of gravitons. However, due to the complexity arising from the inconsistency of the proton's energy tensor components, we introduced an effective mass $m$ to facilitate the calculation of the gravitational form factor.

While our results are promising, there are limitations to consider. The model's reliance on certain approximations, such as the effective mass $m$ in the gravitational form factor calculation, may introduce uncertainties. Future research should aim to refine these approximations and explore the long-term implications of our findings. Additionally, extending this model to study other nucleons and their interactions could provide further insights into the broader field of QCD.

In conclusion, our study demonstrates the potential of the holographic VQCD model to provide a more accurate and comprehensive understanding of nucleon structure. By incorporating key QCD characteristics and considering the quark flavor effect, we have achieved results that are in closer agreement with experimental data and lattice calculations. These findings not only contribute to the ongoing efforts to understand the fundamental properties of nucleons but also pave the way for future research in this area.

\section*{Acknowledgments}

We thank Hai-cang Ren and  Yan-Qing Zhao for useful discussions. This work is supported in part by the National Key Research and Development Program of China under Contract No. 2022YFA1604900. This work is also partly supported by the National Natural Science Foundation of China(NSFC) under Grants No. 12435009, and No. 12275104.

\section*{References}

\bibliography{ref}
\end{document}